\documentclass[aps,prl,twocolumn,amsmath,amssymb]{revtex4-2}
\usepackage{graphicx}
\usepackage{amsmath} 
\usepackage{amssymb} 
\usepackage{hyperref} 
\usepackage{xcolor} 
\usepackage{float}
\usepackage{comment}
\usepackage{dcolumn}
\usepackage{bm}
\usepackage[utf8]{inputenc}
\DeclareUnicodeCharacter{2008}{ } 

\begin{document}

\preprint{APS/123-QED}

 \title{Collisionless Bulk Electron Heating in Resonant Low-Voltage Capacitively Coupled Plasmas}

\author{Sarveshwar Sharma$^{1,2}$}
\email{sarvesh@ipr.res.in;sarvsarvesh@gmail.com} 
\author{Igor D. Kaganovich$^3$}
\author{Animesh Kuley$^4$}
\author{Sudip Sengupta$^{1,2}$}
\author{Alexander V. Khrabrov$^3$}
\author{Bhooshan Paradkar$^5$}

\affiliation{$^1$Institute for Plasma Research, Bhat, Gandhinagar, Gujarat 382428, India}
\affiliation{$^2$Homi Bhabha National Institute, Training School Complex, Anushaktinagar, Mumbai 400094, India}
\affiliation{$^3$Princeton Plasma Physics Laboratory, Princeton, New Jersey 08543, USA}
\affiliation{$^4$Department of Physics, Indian Institute of Science, Bangalore 560012, India}
\affiliation{$^5$School of Physical Sciences, UM-DAE Centre for Excellence in Basic Sciences, University of Mumbai, Mumbai 400098, India}

\date{\today}
\begin{abstract}
We investigate collisionless power absorption in resonant, low-pressure capacitively coupled plasmas (CCPs). In these radio-frequency (RF) discharges, the sheath capacitance almost exactly balances the plasma inductance, driving the total RF discharge voltage down to just a few volts. However, plasma persists not only in this ultra–low-voltage regime; it also generates ions that strike the electrodes with kinetic energies substantially exceeding the amplitude of the applied RF voltage. This counterintuitive behavior arises from the presence of a pronounced electrostatic potential well of approximately 40 V within the plasma bulk, which confines electrons while simultaneously accelerating ions toward the electrodes. We show that, under these resonant conditions, collisionless electron heating exhibits a fundamentally different behavior from the conventional paradigm of stochastic sheath heating mediated by electron–sheath interactions. Instead, the predominant energy transfer mechanism is bulk electron heating in RF electric fields via a primarily collisionless process that emerges from the synergistic action of: (i) a strongly amplified RF electric field within the plasma bulk, (ii) electron oscillatory motion (bouncing) within the plasma potential well, and (iii) electron scattering resulting from collisions with neutral atoms. Collectively, these phenomena give rise to a pronounced high-energy tail in the electron energy distribution function and thereby lead to a substantial enhancement of the ionization rates. As the gas pressure rises, the resonance is disrupted. At the same time, the region of maximum power absorption moves from the plasma core toward the edges and the sheath, which is accompanied by the disappearance of the high‑energy electron population and a corresponding decrease in ionization rates.
\end{abstract}

\maketitle
Capacitively coupled plasma (CCP) discharges underpin modern plasma technologies and semiconductor fabrication. Recently, higher frequency CCPs (30–300 MHz) gained attention for offering higher plasma density, better uniformity, reduced DC bias. Operating at very low pressures in the mTorr range and using low voltages enables finer control of ion and radical fluxes, which is essential for atomic-scale precision. However, sustaining a plasma under such low-voltage, low-pressure conditions is challenging because it functions close to the threshold at which it can be maintained. Under these operating conditions, the electron mean free path exceeds the electrode gap and power dissipation is collisionless. 
Experimental studies have demonstrated that plasma can be sustained in mercury at pressures as low as 0.19 mTorr with an applied RF voltage of only 4 V at 40.8 MHz \cite{Popov_53_1985, Godyak_5_227_1979}, and preliminary simulation investigations of electron series resonance have also been reported in Ref. \cite{Qiu_12_57_PSST_2003}

Plasma sustainment under very low-pressure and low-voltage conditions critically depends on the presence of a high-energy tail in the electron energy distribution, enabling a subset of electrons to exceed the ionization threshold. In addition, it relies on the establishment of self-consistent electric fields that reflect electrons with energies below the ionization potential, thereby confining them within the plasma. This confinement facilitates repeated interactions with the electric field, allowing these electrons to be heated and eventually contribute to enhanced ionization. Nevertheless, the specific mechanisms governing their trapping and their quantitative contribution to ionization processes and collisionless heating remain not adequately characterized and have not yet been comprehensively addressed in the existing literature \cite{Makabe_2006,PhysRevLett.133.235302,PhysRevLett.134.045301,PhysRevLett.114.125002,PhysRevLett.122.185002,PhysRevLett.89.265006,d51374fc,PhysRevLett.93.085003,Turner_1312_1995, Gozadinos_135004_2001}. A detailed understanding of low-voltage, low-pressure plasma discharges is essential for advanced plasma-processing technologies, particularly plasma-assisted Atomic Layer Etching (ALE), where material is removed with near atomic-scale precision through cyclic surface modification and etching \cite{Lemme_NC_13_2022,Akinwande_Nature_573_2019,Das_NatureElec_4_2021,Quellmalz_NatureComminication_12_2021,Joseph_JPCB_129_2025,Kanarik_JPCL_9_2018,Sang_JPDAP_53_2020, Arts_PSST_31_2022, Kim_EM_19_2023, Fischer_POP_30_2023}. Operating with low-energy ions below the physical sputtering threshold minimizes subsurface damage, making such discharges highly attractive for ALE, monolayer removal, smoothing of two-dimensional materials, and the controlled fabrication of surface-sensitive structures such as near-surface nitrogen-vacancy (NV) centers in single-crystal diamond \cite{Walton_JVSTA_39_2021, Jack_JPDAP_58_2025, Shibanov_Vacuum_231_2025, Oscar_JVSTA_43_2025, Draney_JVSTA_44_2026, Toros_DRM_108_2020}.  \par
In this study, we investigate the physics of the low voltage low pressure (LVLP) discharge regime using particle-in-cell simulations with a well-validated \cite{Turner_2013} and widely used \cite{Sharma_110701_2016, Sharma_080705_2018, Sharma_365201_2019, Sharma_063501_2018, Sharma_103508_2019, Sharma_045003_2020, Sharma_103502_2021, Sharma_073506_2023, Turner_2069_1996, Boyle_493_2004, Lauro-Taroni_2216_2004, Sharma_055205_2021, Sharma_275202_2022} code. Plasma breakdown is difficult to initiate under these conditions; however, the LVLP regime can be reached by first igniting the plasma with increased RF power or voltage and then gradually lowering these settings. To model this process, we initialize the simulations with an initial plasma density of $5\times 10^{14}$  $m^{-3}$, assumed to be produced by higher RF coupling that facilitates the breakdown. Maintenance of discharge at low voltage levels is then achieved by applying a $5$ V, $60$ MHz RF voltage to the powered electrode throughout the simulation period. The simulations are performed in a voltage-driven mode (without an external capacitor) with an applied sinusoidal voltage waveform in the form of
\begin{equation}
V_{rf}(t) = V_0 \sin{\left(\omega_{rf} t+\phi\right)}.
\end{equation}
Here, applied voltage ($V_0$), RF frequency ($\omega_{rf}$) are set to $5$ V and $60$ MHz, respectively. We simulate the dynamics of an argon plasma, in the pressure range of $2$ to $30$ mTorr, with an electrode gap of $32$ mm. 
The simulation considers ion-neutral (elastic, inelastic, charge exchange) and electron-neutral (elastic, inelastic, ionization) interactions. It accounts for two metastable argon states, $Ar^*$ ($11.6$ eV) and $Ar^{⁎⁎}$ ($13.1$ eV), along with processes like de-excitation, superelastic collisions, metastable pooling, and multi-step ionizations. Collision cross-section data are sourced from vetted references \cite{Lauro-Taroni_2216_2004, Rauf_2805_1997}. Perfectly absorbing electrodes do not produce secondary electron emission (SEE), which are insignificant when both the pressure and the voltage are low. The detailed reaction set is available in Ref. \cite{Sharma_063501_2018}. The simulations were carried out for more $5000$ RF cycles to achieve steady-state. Stability and accuracy were ensured by selecting a suitable grid size ($\Delta x$) and time step to resolve the Debye length ($\lambda_{De}$) and electron plasma frequency ($\omega_{pe}$) \cite{Birdsall_2004}. \par
At the initial stage, the low applied voltage leads to substantial plasma losses owing to the weak sheath electric field, which is insufficient to effectively confine electrons. As the discharge evolves, the plasma potential increases to mitigate these losses for enhancing electron confinement. Eventually, a steady-state is established in which the plasma potential attains a value considerably higher than the applied RF voltage. For example, in simulations conducted with an applied voltage of $5$ V and a neutral gas pressure of $2$ mTorr, the maximum plasma potential rises nearly ten times the applied voltage. This pronounced increase in plasma potential is indicative of significant bulk electron heating under steady-state operating conditions.. \par

\begin{figure}
    \centering
    \includegraphics[width=0.5\textwidth]{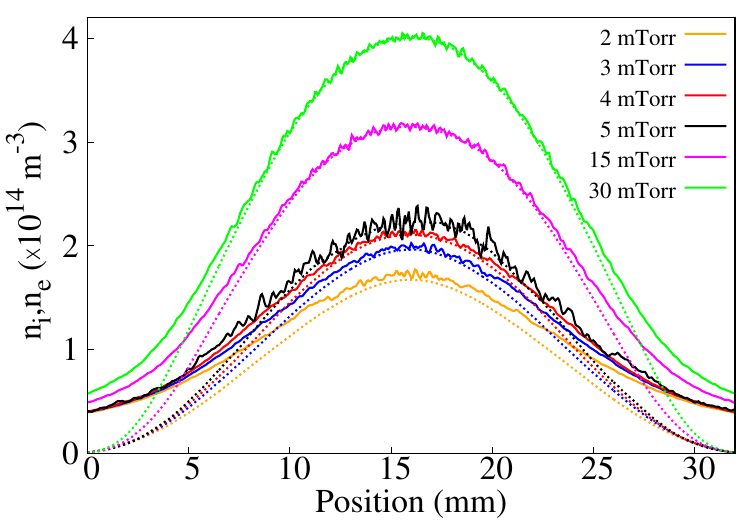} 
    \caption{ Electron and ion density profiles for pressures of $2,3,4,5,15$ and $30$ mTorrs.
}
    \label{Figure5}
\end{figure}
\begin{figure}
    \centering
    \includegraphics[width=0.45\textwidth]{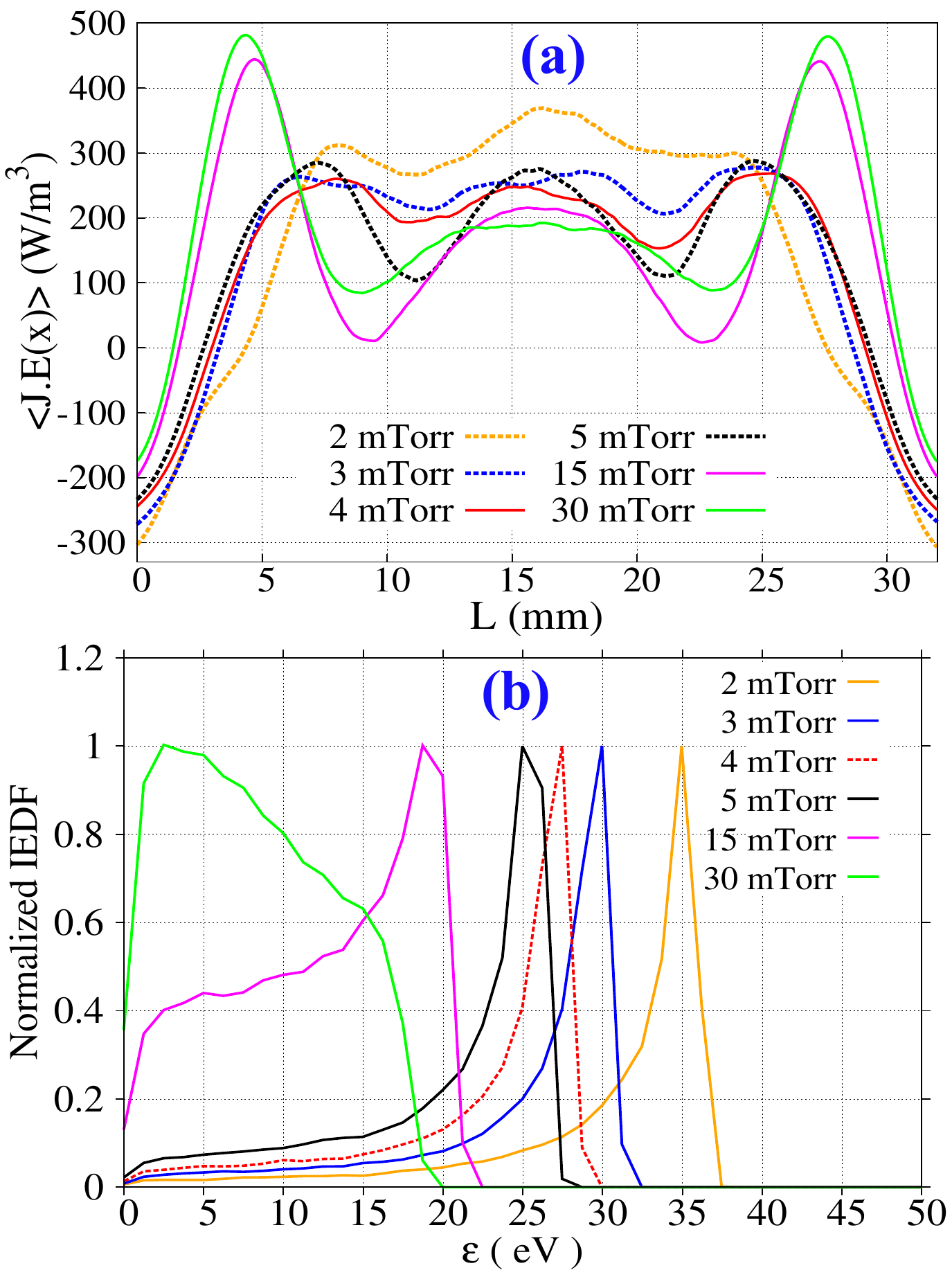} 
    \caption{(a) Time-average electron power absorption (b) IEDF at the powered electrode, for the range of pressure from $2$ mTorr to $30$ mTorr. Operating parameters: $V_0$ $=$ $5$ V, $f_{rf}$ $=$ $60$ MHz.
}
    \label{Figure1_final}
\end{figure}

The steady state profiles of electron and ion densities and the absorption of electron power averaged in time (over $1000$ RF cycles) $\langle J.E(x) \rangle$, in a pressure range of $2$ mTorr to $30$ mTorr, are shown in Figure  \ref{Figure5} and Fig. ~\ref{Figure1_final}(a).
We can see that plasma density in this regime is low and its value is determined by the series resonance condition: that is, the sheath capacitance is balanced by plasma inductance, or $\omega=\omega_{pe} \sqrt{S/L}$, where  \(\omega_{pe}\) is the plasma electron frequency, \(S\) the mean sheath thickness and L the effective bulk length between the electrodes. For the 60 MHz operating frequency, the plasma resonance condition, $\omega=\omega_{pe} $, occurs for the density \(n_{pr}=0.45\times 10^8cm^{-3}\); for the $L/S \simeq 5-6$ the plasma series resonance occurs for the density $ (5-6) \times n_{pr}$ corresponding to the densities values in \ref{Figure5}. For higher pressures, the sheath is significantly narrow, resulting in a significantly higher density at these pressures. The low pressure regime of $2$ mTorr is drastically different from the higher pressure regimes, and we focus on the reasons for the different properties of the discharge compared to the conventional CCP discharges. The first to notice is that at the lowest pressure of $2$ mTorr, power absorption is highest in the bulk plasma region ($\sim 8-25$ mm), which contrasts sharply with conventional CCP behavior where electron heating is mainly confined near the sheath edges \cite{Kawamura_053506_2006, Sharma_110701_2016, Wilczek_125010_2018, Sharma_103508_2019, Lieberman_2005}. Additionally, we see a significant enhancement in the ion energies at low pressure discharges (see Figure ~\ref{Figure1_final}(b)). In low pressure discharge at $2$ mTorr, we observe the maximum ion energy of approximately $35$ eV with quasi-monoenergetic energy spectrum. This marks a significant departure from the broad energy spectrum of ion energies in high pressure discharge at $30$ mTorr having a cut-off energy of about $20$ eV. These results show that operating at very low pressure provides clear benefits: precise control on ion acceleration with a mono-energetic ion distribution functions. Indeed, these properties of the ion spectrum are desirable in the precision etching applications in the semiconductor industry. \par
\begin{figure}[htbp]
    \centering
    \includegraphics[width=0.5\textwidth]{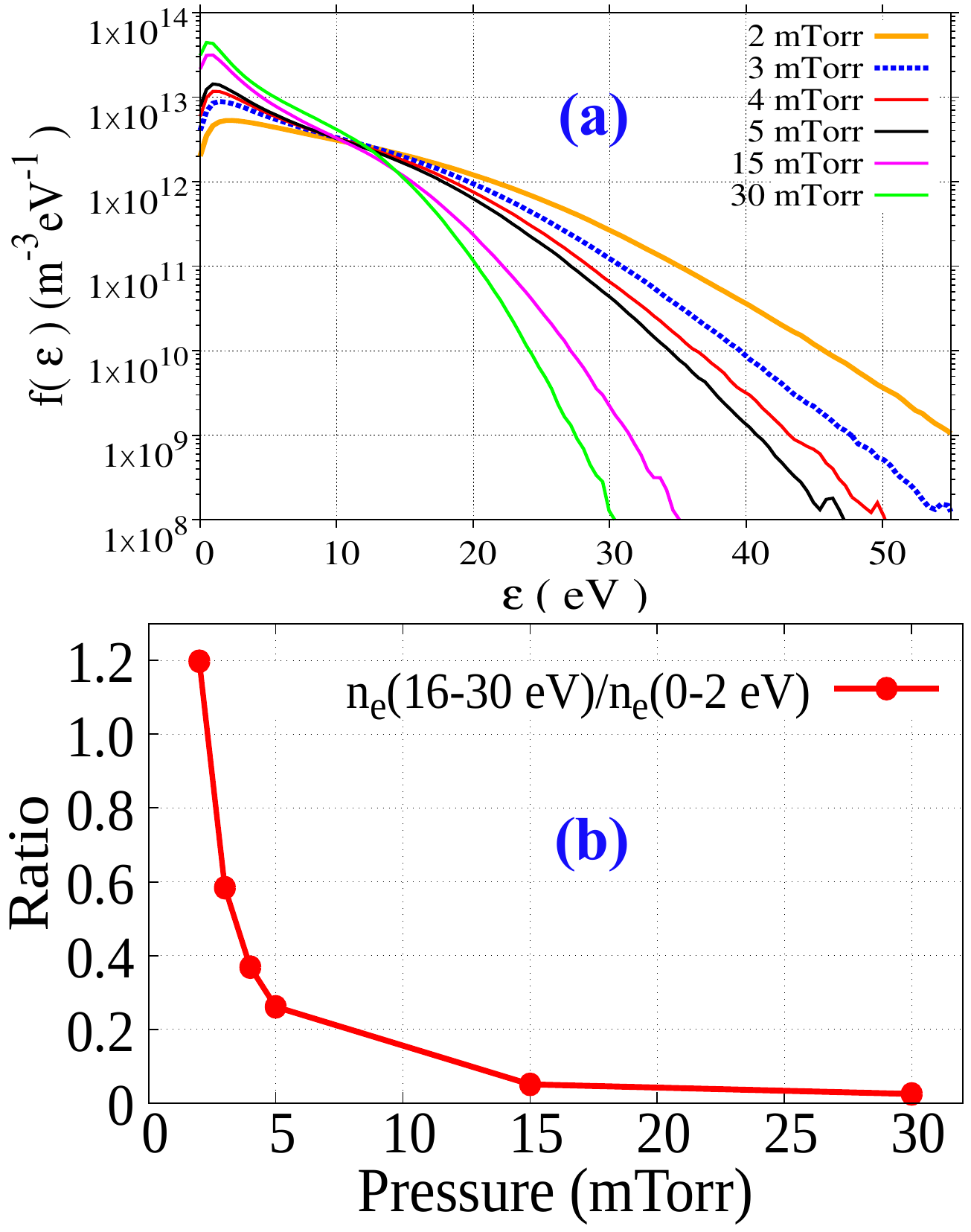} 
    \caption{The figure (a) EEDF at the center of the discharge ranging from $2$ mTorr to $30$ mTorr. (b) presents the pressure-dependent variation in the ratio of high-energy ($16–30$ eV) to low-energy ($0–2$ eV) electron densities, offering insights into how energetic electron populations evolve with pressure.}
    \label{Figure2_final}
\end{figure}

The electron energy distribution function (EEDF), shown in Figure ~\ref{Figure2_final}(a), indicates that the fraction of high-energy non-thermal electrons increases as the operating pressure decreases. A pronounced transition, manifested as a knee in the EEDF in the vicinity of the first excited state of argon (11.6 eV), is observed consistently across all measured energy spectra. At $2$ mTorr, the Druyvesteyn-like EEDF \cite{Godyak_233001_2011} exhibits a prominent high-energy tail, with a substantial population of electrons possessing energies exceeding $60$ eV. The relative contribution of non-thermal to thermal electrons, estimated as a ratio of number of high energy electrons ($16 - 30$ eV) to thermal electrons ($0-2$ eV),  for varying operating pressures is shown in Figure ~\ref{Figure2_final}(b). This result indicates that at a pressure of \(2\) mTorr, the population of non-thermal electrons exceeds that of thermal electrons. However, as pressure increases, the relative contribution of non-thermal electrons progressively decreases, becoming negligible at \(30\) mTorr, in agreement with the behavior typically observed in conventional low-pressure discharges. For example, even at $5$ mTorr operating pressure, typically considered as the low pressure CCP, the contribution of high energy electrons is about $20\%$ of the thermal electrons. On the other hand, for the $2$ mTorr case, we find that the number of high energy electrons have exceeded significantly (around $120\%$) the thermal electrons. Note that this behavior is fundamentally different from the standard low pressure discharge, typically encountered in CCPs.  
 This tendency can be understood by estimating the mean free path of the electron $\lambda_{mfp}$ using the electron–neutral collision cross section ($\sim 1.5\times 10^{-19}$ $m^2$) at various operating pressures. Typical values of $\lambda_{mfp}$ in terms of system size ($L = 32$ mm) for various operating pressures are shown in Table 1.
\begin{table}[htbp]
\centering
\begin{tabular}{|c|c|c|c|c|}
\hline
Pressure & 2 mTorr & 5 mTorr & 15 mTorr & 30 mTorr \\
\hline
$\lambda_{mfp}/L$ & 3.2 & 1.2 & 0.6 & 0.2 \\
\hline
\end{tabular}
\caption{Ratio of the electron mean free path ($\lambda_{mfp}$) to the electrode gap length ($L$) at various operating pressures. 
}
\label{tab:mfp_ratio}
\end{table}

This clearly shows that at operating pressures below $5$ mTorr, electron heating is governed predominantly by collisionless processes, in which electrons gain energy through repeated interactions with the oscillating bulk electric fields. 

\begin{figure}
    \centering
    \includegraphics[width=0.5\textwidth]{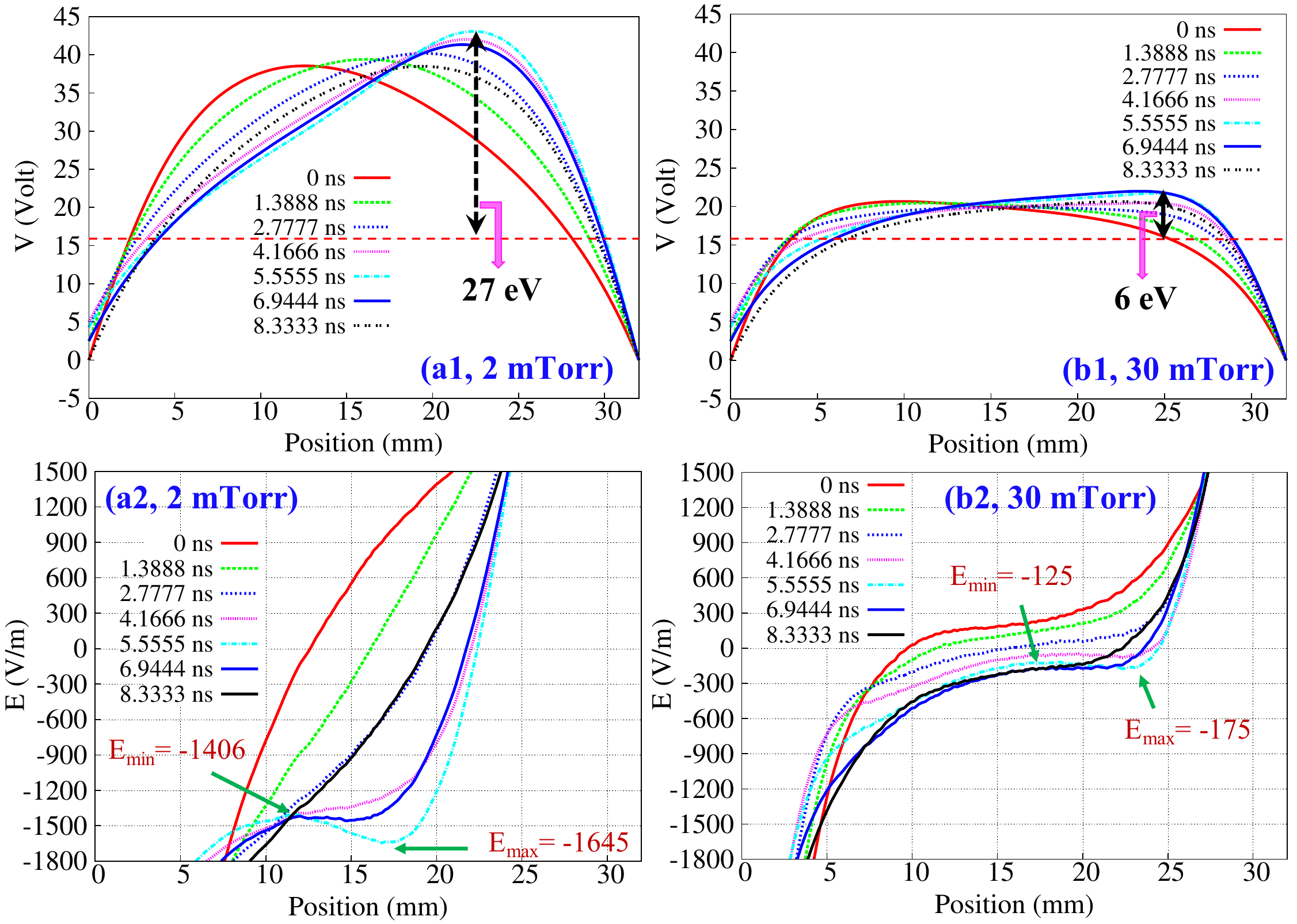} 
    \caption{The figure presents snapshots of the potential and electric field (zoomed-in) profiles for $2$ mTorr (a1, a2) and $30$ mTorr (b1, b2), respectively. Only the first half of the RF cycle is shown here, as the second half exhibits a symmetrical repetition of the same pattern.
}
    \label{Figure3_final}
\end{figure}

To elucidate the mechanisms of electron heating in greater detail, we examined the temporal evolution of the electrostatic potential and associated electric fields over the duration of a single RF cycle. Figure ~\ref{Figure3_final} shows the snapshots of the potential and electric field (zoomed-in) profiles for $2$ mTorr (a1, a2) and $30$ mTorr (b1, b2), respectively. Only the first half of the RF cycle is shown here, as the second half exhibits a symmetrical repetition of the same pattern. Figures ~\ref{Figure3_final} (a1) correspond to the potential profile at $2$ mTorr where distinct potential structures form in specific phases of RF, notably at $5.55$ ns. A dashed red line is also plotted which corresponds to the ionization potential of argon (i.e. $15.76$ eV) in the figure (a1). The magnitude of the potential well in the plasma bulk is observed to be as high as $27$ eV, which is high enough to trap the energetic electron in the plasma bulk. These energetic electrons heated predominately by the bulk RF electric field can produce ionization and maintain the discharge without interaction with the sheathes, as conventionally considered to be the case for CCP.  


Figure ~\ref{Figure3_final} (a2) presents snapshots of the electric field profile, for the same phases of the RF cycle as for the potential profiles shown in Fig. \ref{Figure3_final} (a1). The y-scale is limited to focus on the bulk region because the electric field is significantly stronger inside the sheath. At $2$ mTorr, a non-monotonic profile of the electric field forms at $5.55$ ns.\\
When the pressure increases to $30$ mTorr, the overall potential profile decreases, accompanied by a pronounced reduction in the amplitude of the electric field, as illustrated in Fig. ~\ref{Figure3_final} (b2). At this pressure, the potential well in the bulk (represented by the dashed red line) is limited to approximately $6$ eV. This magnitude is insufficient to effectively confine energetic electrons in the plasma bulk. Correspondingly, heating in interactions with the RF sheaths becomes dominant, as is evident in Fig. \ref{Figure1_final}(a).\\
\begin{figure}
    \centering
    \includegraphics[width=0.5\textwidth]{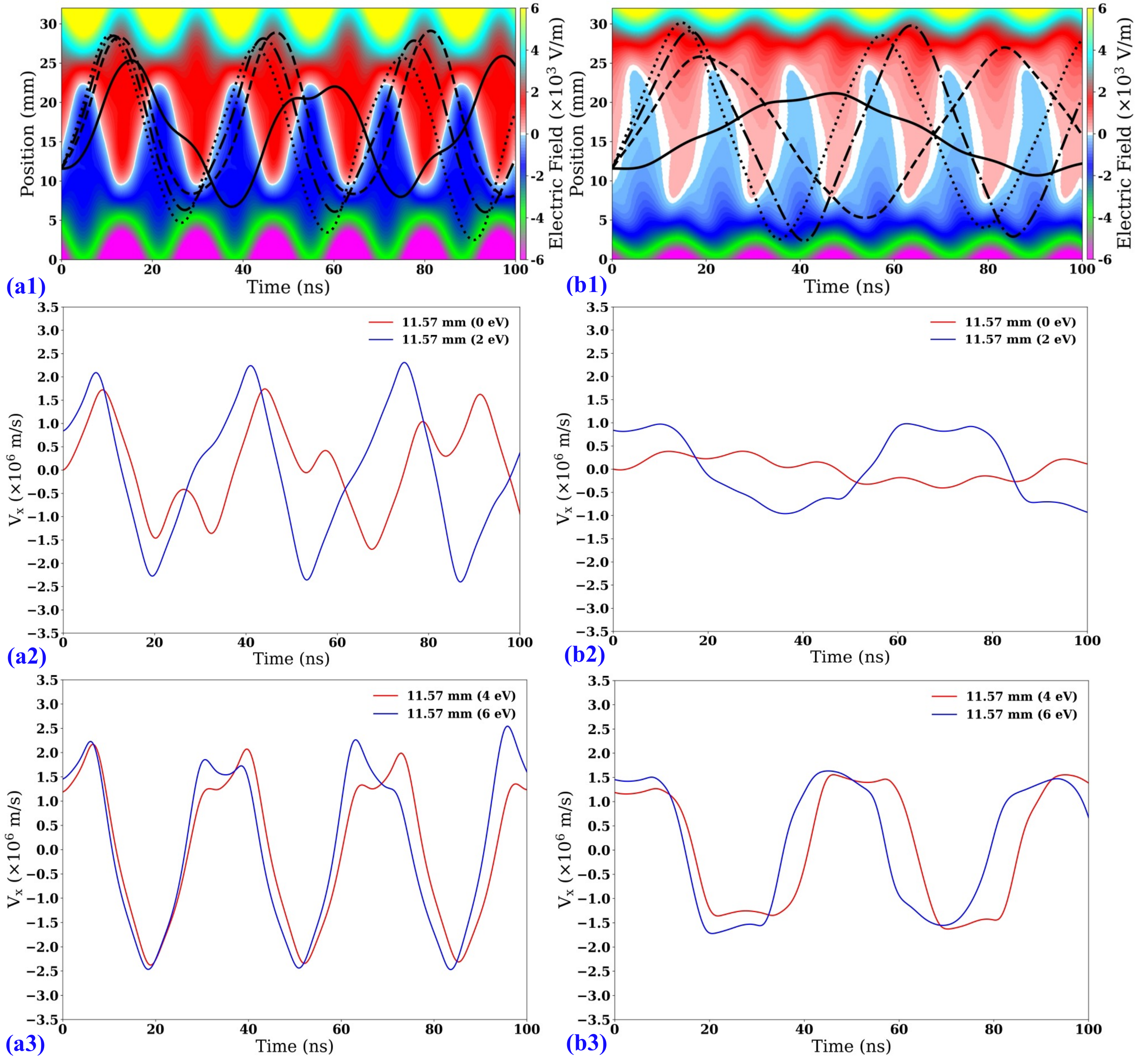} 
    \caption{ Spatiotemporal distributions of the electric field over six consecutive RF cycles with superimposed trajectories of injected electrons at $x=11.57$ mm with energies trajectories  for initial energies $0,2,4,6$ eV for pressures of $2$ mTorr (a1) and $30$ mTorr (b1), respectively. Panels (a2, a3) and (b2, b3) depict the corresponding trajectories  for initial energies $0,2$ and $4,6$ eV, respectively. These comparisons highlight the influence of pressure on both the field dynamics and the mechanisms of electron heating.
}
    \label{FigureTrajectory}
\end{figure}
At the low-pressure condition of $2$ mTorr, Figure ~\ref{FigureTrajectory} (a1) shows that the significant oscillating electric fields appear in the center of the discharge' (visually represented by the alternating red and blue zones). As pressure increases, a significant reduction of the electric field in the plasma bulk is observed, as shown in Fig. ~\ref{FigureTrajectory} (b1).
To demonstrate the difference, the electron trajectories are also shown in Figs. ~\ref{FigureTrajectory} (a2, a3) for $2$ mTorr and Figs. ~\ref{FigureTrajectory} (b2, b3) for $30$ mTorr. First, note that electrons with initial energies of \(0, 2\ \text{eV}\) are confined within the plasma bulk in the $30$ mTorr case, whereas they are not confined in the $2$ mTorr case. This behavior arises because, at $30$ mTorr, relatively weak RF electric fields do not accelerate electrons to energies that exceed the confining potential well, while at $2$ mTorr the RF fields are sufficiently strong to accelerate electrons beyond the confining potential barrier. Second, note the difference for the electron trajectories for the cases $2$ mTorr and $30$ mTorr for electrons with initial energies $4,6$ eV shown in Figs.~\ref{FigureTrajectory} (a3, b3). For $30$ mTorr, the electron energy changes mainly in interaction with the sheath, while interaction with the bulk electric field results in insignificant changes in the electron energy, as evident by straight trajectories in the bulk with approximately constant velocity. The opposite is true for $2$ mTorr, a significant electron pickup energy in the bulk electric field; the electron velocity variations in the bulk and in interactions with the sheath are comparable. The analytical theory for such heating was proposed on the basis of separation of the RF electric and ambipolar confining electric field in \cite{Aliev_1997} assuming that the RF field is small compared to the ambipolar field. This assumption is satisfied for the $30$ mTorr case but is invalid for the $2$ mTorr case. However, qualitative conclusions can still be applied that the sheath and the bulk field counteract each other as is evident by examining the trajectory in Fig.~\ref{FigureTrajectory} (a3).  



In conclusion, LVLP CCP discharges were investigated using 1D-3V PIC/MCC simulations to examine kinetic effects at sub$-$5$-$mTorr pressures. In this regime, bulk collisionless electron heating dominates over traditional electron heating in interactions with the RF sheath fields, producing high-energy electrons that sustain ionization despite low neutral densities. The plasma potential well forms much above the applied 5 V RF voltage confining energetic electrons. At 2 mTorr, the ion energy distribution becomes quasi-monoenergetic ($\sim$35 eV), in contrast to the broader ion energy with a maximum at $\sim$20 eV at 30 mTorr due to the shorter mean free path at these pressures. This regime of collisionless heating is significantly different from the standard low pressure regime where electron heating is predominantly due to interaction with the sheath electric field and power dissipation is peaked near sheath edges. In contrast, the LVLP regime is characterized by bulk heating through in enhanced RF electric field in the plasma bulk. This study opens new opportunities in plasma processing that utilize these distinct characteristics of the LVLP regime.\\
\textbf{Acknowledgments:} IDK’s research was supported by the US DOE, Office of Fusion Energy Science under the DE-AC02-09CH11466 contract, as part of the Princeton Collaborative Low Temperature Plasma Research Facility (PCRF).  AK and SS  work is supported by the Board of Research in Nuclear Sciences (BRNS) and the ARG program of the Anusandhan National Research Foundation (ANRF). SS thanks Prof. Miles M. Turner for the original PIC code.

\begin{figure}
    \centering
    \includegraphics[width=0.5\textwidth]{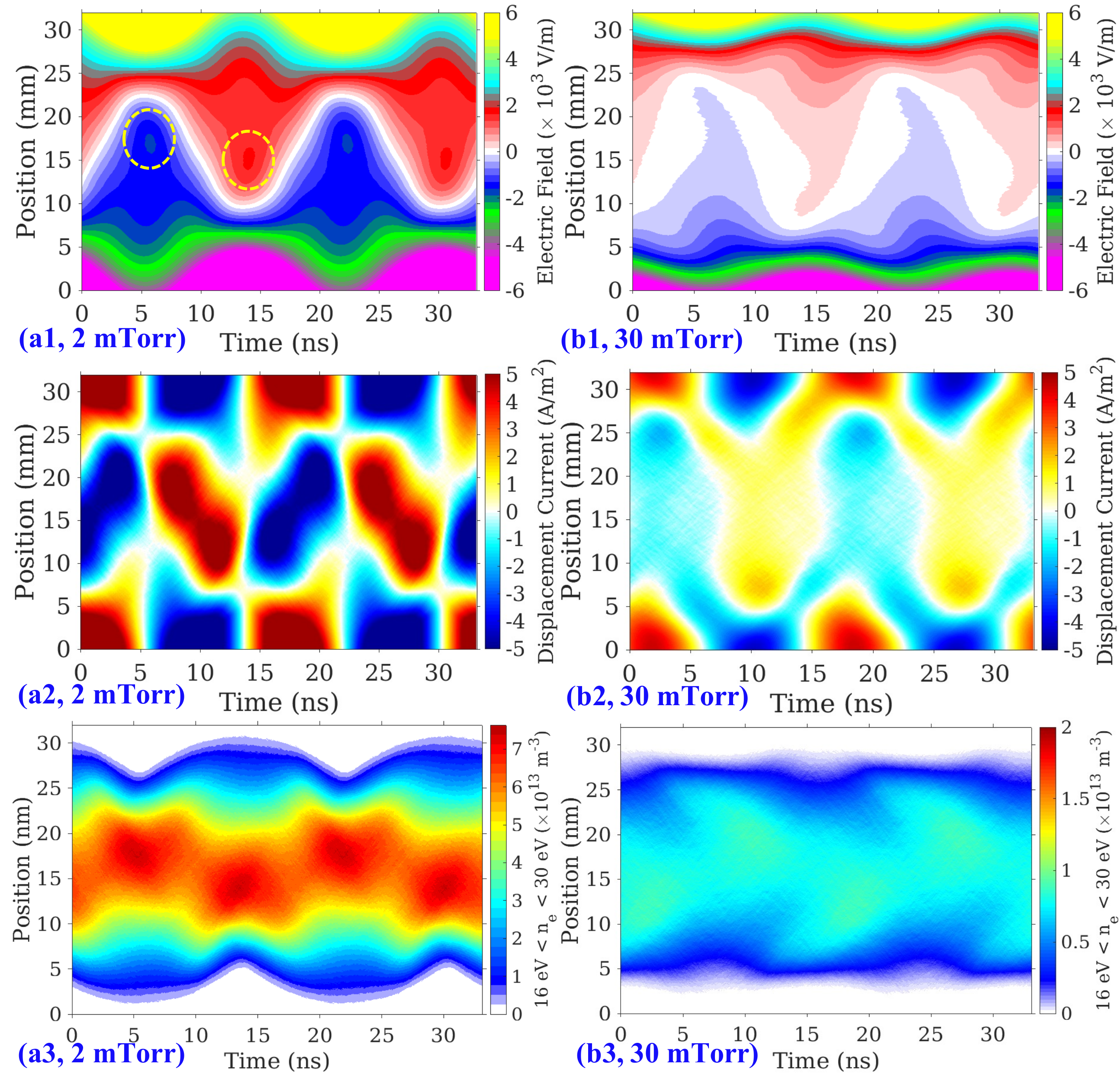} 
    \caption{ Spatiotemporal distributions of the electric field and displacement current over two consecutive RF cycles are illustrated for pressures of $2$ mTorr (a1, a2) and $30$ mTorr (b1, b2), respectively. Panels (a3) and (b3) depict the corresponding electron density distributions within the energy range of $16-30$ eV at $2$ mTorr and $30$ mTorr, respectively. These comparisons highlight the influence of pressure on both the field dynamics and the population of high-energy electrons.
}
    \label{Figure4_final}
\end{figure}

\textbf{End matter:}
 Figure ~\ref{Figure4_final} shows  the electric field in the plasma bulk and emphasizes the difference between $2$  and $30$ mTorrs. For  $30$ mTorr, Fig. ~\ref{Figure4_final}  (b1), the electric field in the plasma bulk is predominantly localized within narrow spatial regions located adjacent to the electrode sheath boundaries.  Because the electric field decays more rapidly away from the sheath, collisionless heating of bulk electrons is greatly reduced at these pressures. This transition clearly marks the shift from a collisionless regime in the bulk, where energetic electrons can interact with strong electric fields in the bulk, to a more conventional regime where energy transfer is dominated by interactions with the sheath fields. \par
At $2$ mTorr (Figure ~\ref{Figure4_final} (a2)), the displacement current is not only intense, but also extends significantly into the plasma bulk. The magnitude of the displacement current in these bulk regions ranges between $6–7$ $A/m^2$, with the temporal evolution of the peak-to-peak current at the discharge center reaching approximately $12$ $A/m^2$. This indicates that the plasma frequency is comparable to the discharge frequency. In contrast, at $30$ mTorr (Figure ~\ref{Figure4_final} (b2)), the displacement current ($ \sim 2$ $A/m^2$) is considerably reduced within the bulk and is primarily concentrated in the sheath regions. 
Figures ~\ref{Figure4_final} (a3) and (b3) show the spatiotemporal density of electrons in the $16-30$ eV energy range at 2 mTorr and 30 mTorr, respectively. Since the ionization potential of argon is 15.76 eV, these energetic electrons play a key role in sustaining ionization. At 2 mTorr, the trapped-electron regions are well localized and exhibit high density, whereas at 30 mTorr they become more diffuse and significantly weaker. The trapped-electron density decreases from $\sim 7.6\times 10^{13}$ $m^{-3}$ at 2 mTorr to $\sim 1\times 10^{13}$ $m^{-3}$ at 30 mTorr, a reduction of more than seven times.
\bibliographystyle{apsrev4-2}
\bibliography{Ref_LVLP}


\end{document}